\documentclass{article}
\usepackage[utf8]{inputenc}
\usepackage[T1]{fontenc}
\usepackage{settings}

\newcommand{\red}[1]{\textcolor{red}{#1}}

\begin{document}
\captionsetup{margin=10pt,font=footnotesize,labelfont=bf,labelsep=endash,justification=centerlast}

\begin{center}
    \LARGE
    \textbf{Agent-based Integrated Assessment Models: Alternative Foundations to the Environment-Energy-Economics Nexus}\\[4ex]
\end{center}

\begin{center}
\small
\noindent \textbf{Karl Naumann-Woleske} (0000-0001-8182-9256)\\
Chair of EconophysiX and Complex Systems, Laboratoire d'Hydrodynamique, CNRS, Ecole Polytechynique, Institut Polytechnique de Paris, 91120 Palaiseau, France

\href{mailto:karl.naumann@ladhyx.polytechnique.fr}{karl.naumann@ladhyx.polytechnique.fr}\\[2ex]
\end{center}

\vfill
\begin{center}
    \textbf{Abstract}
\end{center}
Climate change is a major global challenge today. To assess how policies may lead to mitigation, economists have developed Integrated Assessment Models, however, most of the equilibrium based models have faced heavy critiques. Agent-based models have recently come to the fore as an alternative macroeconomic modeling framework. In this paper, four Agent-based Integrated Assessment Models linking environment, energy and economy are reviewed. These models have several advantages over existing models in terms of their heterogeneous agents, the allocation of damages amongst the individual agents, representation of the financial system, and policy mixes. While Agent-based Integrated Assessment Models have made strong advances, there are several avenues into which research should be continued, including incorporation of natural resources and spatial dynamics, closer analysis of distributional effects and feedbacks, and multi-sectoral firm network structures. 
\vfill
\newpage

\section{Introduction}

Anthropogenic climate change is one of the major global challenges we face as a society today, with widespread environmental, social and economic effects \citep{IPCC2022ClimateChange2022}.
To assess the economic impact of climate change and develop economic policies, economists have developed Integrated Assessment Models (IAMs) that link economic dynamics with environmental aspects.
These models are central in assessments of climate change mitigation strategies and their implications \citep[e.g. see][]{IPCC2018GlobalWarmingIPCC}.
However, existing models have been subject to strong critique.
Agent-based Models have recently come to the fore as an alternative framework for macroeconomic modelling, as well as environment-energy-economics modeling \citep{BalintEtAl2017ComplexityEconomicsClimate, CiarliSavona2019ModellingEvolutionEconomic, LampertiEtAl2019AgentbasedIntegratedAssessment}. 
In this paper, I analyze four Agent-based Integrated Assessment models (ABIAMs) and how they respond to the critiques of the currently mainstream equilibrium-based IAMs.
Previous reviews of the application of complexity economics and agent-based modeling to climate issues \citep[e.g.][]{CastroEtAl2020ReviewAgentbasedModeling, BalintEtAl2017ComplexityEconomicsClimate, LampertiEtAl2019AgentbasedIntegratedAssessment} have focused on the general benefits of these approaches, but there has been no systematic comparison of existing ABIAM models.\footnote{\citet{LampertiEtAl2019AgentbasedIntegratedAssessment} showcased the usefulness of the ABIAM approach by highlighting the work of \citet{LampertiEtAl2018FarawayCloseCoupled} and \citet{WolfEtAl2013MultiagentModelSeveral}, but did not compare models in detail.}

For the purposes of this paper, we consider an IAM to be a model with endogenous and linked climate and economic modules.\footnote{This is a rather restricted definition as there is a wide scope of these models \citep{KreyEtAl2019LookingHoodComparison}. However it is appropriate for this paper, as we are addressing economy-environment linkages and representations of the economic system here. For this paper, we consider primarily the cost-benefit type IAMs following the Dynamic Integrated Climate Economics (DICE) models of \citet{Nordhaus1992OptimalTransitionPath} and the process-driven IAMs used in \citet{IPCC2014ClimateChange2014, IPCC2018GlobalWarmingIPCC, IPCC2022ClimateChange2022}}
The majority of the IAMs meeting this criteria consider a first-best economic system grounded in equilibrium (general or partial) and perfect foresight \citep{ForsterEtAl2018MitigationPathwaysCompatible, KeppoEtAl2021ExploringPossibilitySpace}.\footnote{A main source of the critique is centered around the IAMs coming from the economics literature, and that are based on the work of \citet{Nordhaus1992OptimalTransitionPath}. The IPCC review models tend to have a higher degree of complexity and detail}
It is these models that have faced strong critique for their grounding in equilibrium, foresight and optimization by representative agents in the macroeconomic modules underlying IAM \citep[see][]{AckermanEtAl2009LimitationsIntegratedAssessment, Pindyck2013ClimateChangePolicy, Pindyck2017UseMisuseModels, Stern2013StructureEconomicModeling, Stern2016EconomicsCurrentClimate, Weitzman2013TailHedgeDiscountingSocial, ReveszEtAl2014GlobalWarmingImprove, FarmerEtAl2015ThirdWaveEconomics}.
A recent review by \citet{KreyEtAl2019LookingHoodComparison} considers several distinct areas of critique:
(a) the absence of heterogeneity of actors and within groups of actors, which is key to societal transitions due to social processes emerging from interactions and coordination (e.g. lifestyle change, political actions). The absence of heterogeneity also strongly limits the analysis of distributional effects despite their importance \citep{DiffenbaughBurke2019GlobalWarmingHas}.
(b) Technology and its diffusion is misrepresented either by being exogenous, or, when endogenous, being too optimistic (or pessimistic) in advances and diffusion \citep[see][]{MercureEtAl2019ModellingInnovationMacroeconomics, GambhirEtAl2019ReviewCriticismsIntegrated}.
(c) A lacking representation of the financial system, though it faces large risks \citep[e.g. see][]{Monasterolo2020ClimateChangeFinancial, MonasteroloEtAl2019UncertaintyClimatePolicies} and at the same time may be a driving force in the green transition \citep[see][]{CampiglioEtAl2018ClimateChangeChallenges, Caldecott2018StrandedAssetsEnvironment}.
(d) Energy-economy feedbacks are not fully represented, with missing energy system, material and economy linkages \citep{PauliukEtAl2017IndustrialEcologyIntegrated} and an unrealistic decoupling of economic growth from energy usage or emissions \citep{NietoEtAl2020MacroeconomicModellingEnergy}.
Adding to this list, IAMs, in particular those based on \citet{Nordhaus1992OptimalTransitionPath}, are frequently criticised for their "ad-hoc" \citep{LampertiEtAl2018FarawayCloseCoupled} representation of damages from increases in the temperature anomaly over pre-industrial levels, thus underestimating the costs of climate change and the benefits to a low-carbon economy \citep{Stern2016EconomicsCurrentClimate, Pindyck2017UseMisuseModels}.
In these models, the gradual deterministic reactions to mean surface temperatures omit the emergence of tipping-points, rare events, increasing variability and decreasing predictability of climate conditions \citep{WrightErickson2003IncorporatingCatastrophesIntegrated}. 

The accumulation of concerns and critiques suggests that it might be beneficial to consider alternative frameworks for representing the economy and the environment-economy feedbacks.
Agent-based Integrated Assessment models aim to provide an alternative methodology for modeling the interactions of the socioeconomic system with the biosphere. Interest in the application of Agent-based Modelling to the climate-economy-energy nexus dates back to \citet{MossEtAl2001AgentbasedIntegratedAssessment, Moss2002AgentBasedModelling}, who argued that it can serve as a well validated description of social and natural systems. 
The actual applications of Agent-based Models to the environment-energy-economy nexus has recently been reviewed in \citet{BalintEtAl2017ComplexityEconomicsClimate} and \citet{CastroEtAl2020ReviewAgentbasedModeling}, and includes topics such as  carbon and electricity markets, technology diffusion models, and coalition formation.

Agent-based Models represent complex adaptive systems \citep[see ][]{LeBaronTesfatsion2008ModelingMacroeconomiesOpenEnded, HommesLeBaron2018HandbookComputationalEconomics, KirmanGallegati2022QuarterCenturyComplex}.
Specifically, they comprise a set of heterogeneous agents that interact, adapt and evolve over the course of time according to explicit market protocols.
The macro- and meso features of the model are then obtained by aggregation over the set of all individual agents. 
The interactions give rise to emergent phenomena, such as endogenous crises and oscillations, and can serve as tools to understand out-of-equilibrium dynamics.
The agent-based methodology has already been successfully applied in macroeconomic analysis \citep[see][for a detailed review]{DawidDelliGatti2018ChapterAgentBasedMacroeconomics}.
In relation to the biosphere-economy interaction, ABIAMs offer several distinct advantages such as the built in heterogeneity, a more granular endogenous innovation and diffusion process, a granular representation of the financial system, and agent-specific damage functions.
Their modularity and detail of agent-based models also allows for a closer interaction with stakeholders, such as policymakers. 

The paper proceeds as follows: Section \ref{sec:abiammacro} briefly presents the models under consideration and their macroeconomic backbones, Sections \ref{sec:abiamenergy} and \ref{sec:abiamclimate} compare the energy, resource and climate modules of these ABIAMs. In light of this structure, in Section \ref{sec:abiampolicy} I consider the policy studies and recommendations that these models have been used for. Finally, in Section \ref{sec:nextsteps} I consider several next steps in the ABIAM research stream. Tables \ref{tab:macro}-\ref{tab:policy} in the Appendix give a detailed comparison of the different models in the spirit of \citet{DawidDelliGatti2018ChapterAgentBasedMacroeconomics}, with extensions for the energy (Table \ref{tab:energyresources}) and climate (Table \ref{tab:climate}) modules. These tables may serve as a reference point for the current state of the models going forward.

\section{Agent-based Integrated Assessment Models: Four Candidates}%
\label{sec:abiammacro}
To consider how Agent-based Models may address some of the critiques levied at Integrated Assessment Models, I consider four ABIAM models: the Dystopian Schumpeter meets Keynes model \citep[DSK hereafter,][]{LampertiEtAl2018FarawayCloseCoupled, LampertiEtAl2019PublicCostsClimateinduced, LampertiEtAl2020ClimateChangeGreen, LampertiEtAl2021ThreeGreenFinancial}, the ABMIAM model of \citet{SafarzynskavandenBergh2022ABMIAMOptimalClimate}, the model of \citet{CzuprynaEtAl2020AgentBasedApproachIntegrated} (CFHS hereafter), and the model of \citet{GerdesEtAl2022LaborEnvironmentGlobal} (GRSW hereafter). This section introduces the models, their purposes and general macroeconomic structure. 

The DSK model is a climate-extension of the well-established Keynes+Schumpeter model \citep{DosiEtAl2010SchumpeterMeetingKeynes,DosiEtAl2010SchumpeterMeetingKeynes}.
The purpose of the DSK presented in \citet{LampertiEtAl2018FarawayCloseCoupled, LampertiEtAl2020ClimateChangeGreen} is to provide an alternative to the specification of climate damages due to increases in atmospheric $CO_2$ concentration.
The model consists of heterogeneous firms in the consumption and capital goods sectors, both of which are powered by electricity from an energy sector.
Capital goods firms produce machines, with an innovation and imitation process leading to newer capital vintages (more energy efficient, higher labor productivity) based on R\&D investments.
Consumption good firms invest in different machines to produce a generic consumption good purchased by the households.
These investments are financed by through imperfect capital markets, where firms are subject to credit limitations.
The energy sector produces electricity using stylized green (renewable) and brown (fossil) power plants, which differ in their cost structure and emissions.
Again power plants may invest in R\&D and pursue an imitation-innovation strategy.
Finally, a climate module connects $CO_2$ emissions by all sectors to a carbon cycle, and thus to an increase in the temperature anomaly over pre-industrial times.
This leads to a disaster generating distribution with increasing average damage, and a higher probability of extreme damages to firms' capital stock and labor productivity as $CO_2$ concentration rises.

In \citet{LampertiEtAl2019PublicCostsClimateinduced, LampertiEtAl2021ThreeGreenFinancial}, the DSK model is extended with a detailed set of heterogeneous banks to study the effects of changes in financial regulation on the transition to a low-carbon economy in the presence of financial instability.
Their study responds to a growing body of literature suggesting that climate change feedbacks could be increased by financial market instability, especially with a banking system exposed to physical risks \citep{Monasterolo2020ClimateChangeFinancial}.
In doing so, they also simplify the climate module, thus for the remainder of the paper I will distinguish between the main branch of the DSK model and the extension to a financial system, referred to as DSK-FIN. 

The second model, the ABMIAM proposed by \citet{SafarzynskavandenBergh2022ABMIAMOptimalClimate} studies how heterogeneous agents and emerging inequalities in labor and capital income can lead to larger estimates of the social cost of carbon than the aggregate Dynamic Integrated Climate Economy (DICE) models based on \citet{Nordhaus1992OptimalTransitionPath} that have been subject to strong critique.
Building on the work of \citet{SafarzynskavandenBergh2017FinancialStabilityRisk, SafarzynskavandenBergh2017IntegratedCrisisenergyPolicy}, the ABMIAM considers a set of consumption good firms with quality differentiated products, which households purchase based on a combination of quality, price and peer-effects.
Over time, firms can improve their production technology by investing in R\&D.
To expand, firms may request loans from a bank which is subject to credit regulation and an interbank market.
To produce goods, the firms apply labor, capital and electricity.
The energy sector is based on heterogeneous power plants fueled by gas, coal or renewable energy with different cost structures, efficiencies and emission intensities.
The emissions of the energy sector lead to increases in the atmospheric stock of carbon, and consequent damage feedbacks.
These were modeled like the cost-benefit DICE IAM to compare calculations of the social cost of carbon, but unlike DICE the shocks are distributed amongst agents based on their relative share of wealth similar to \citet{DennigEtAl2015InequalityClimateImpacts}. 

\citet{CzuprynaEtAl2020AgentBasedApproachIntegrated} propose a global ABIAM model including ten regions and multiple different goods.
In each region consumer and capital firms produce their respective goods, and households purchase with preferences for price and a minimum necessary consumption.
To expand their capacities, firms receive investments from households rather than a financial system, paying back a share of profits as returns.
As in the previous models, production requires the use of electricity which is supplied by a two-layer energy sector.
In the first instance, fuel-extraction firms mine from finite regional stocks.
The collected fossil fuels are then sold to a heterogeneous set of power plants, who provide electricity regionally.
Over time, production technology improves at an exogenously given rate, with renewable technology costs also dropping.
The combustion of fossil fuels increases the mean temperature anomaly over pre-industrial times, which leads to region-specific agricultural shocks (reduced productivity), labor productivity reductions, and natural disasters (capital damage).

Finally, \citet{GerdesEtAl2022LaborEnvironmentGlobal} propose a fourth ABIAM model.\footnote{The authors themselves do not consider it to be an IAM, however, they do link the economy and environment both ways, so it falls under the definition of an IAM as applied here}
The GRSW model considers two regions, one with a mining sector (a stylized global south) and one with a capital goods sector (a stylized global north).
Both regions have local labor and consumption good markets, with a lower wage in the south than the north.
Households purchase a homogeneous consumption good from their regional consumption firms, who produce using capital and labor.
The capital goods sector additionally requires resources from the mining sector in the south.
In the GRSW model, the mines cause localized pollution that reduces workers health and thus labor productivity.
Simultaneously, the capital goods sector emits $CO_2$ when producing machines, which causes natural disaster shocks to hit firms across both regions.
Using this model, the authors explore unequal exchange and a double burden faced by the global south (economic dependence on the north, resulting in lower wages, and heavy damage to the local environment).

In terms of the economic systems represented, these four models differ primarily in their representations of the financial system, with more detail in the DSK-FIN and ABMIAM, innovation (endogenous methods in DSK, ABMIAM and GRSW), spatial scope, energy system (see Section \ref{sec:abiamenergy}), and policy institutions (see Section \ref{sec:abiampolicy}). With respect to the climate modules, the structure in each is similar and oriented to $CO_2$ emissions, they differ in the types of damages andc their distribution (Section \ref{sec:abiamclimate}).  

By construction, each of the four main models considered in this review respond to the IAM critique of missing heterogeneity and interactions.
Even when initialized with perfectly similar agents, over time the interactions and market protocols lead to emerging heterogeneity, such as through differentiated innovation.
Many of the models, such as the ABMIAM or CFHS initially impose some form of heterogeneity in order to study its effects and to calibrate the model more closely to empirically observed heterogeneity.
The results indicate that this heterogeneity is indeed important.
For example, \citet{SafarzynskavandenBergh2022ABMIAMOptimalClimate} find that reducing inequalities in labor income leads to a larger social cost of carbon, while inequalities in capital income (rent) lead to a larger share of the population driven into poverty, thus reducing GDP and emissions, and consequently the social cost of carbon.

In terms of geographic and spatial heterogeneity, the considered ABIAMs lag behind their process-based counterparts.
Only the CFHS model represents multiple consumption sectors and a larger number of regions, though both are still small in comparison to some other more heterodox IAMs such as ACCLIMATE \citep{OttoEtAl2017ModelingLossPropagationGlobal} or E3ME \citep{MercureEtAl2018EnvironmentalImpactAssessment}.
Incorporating this might lead to a more detailed description of the production process and the global value chains that underlie it.
Agent-based Models have already been applied successfully to production networks, and could be integrated here as well \citep[e.g. see the models of][]{WolfEtAl2013MultiagentModelSeveral,GualdiMandel2019EndogenousGrowthProduction}.

Across the DSK, ABIAM and GRSW, innovation and technology diffusion are also endogenized.
For instance, through the innovation-imitation process of different firms that lead to new technologies but also to the imitation of technology that are in use by industry leaders.
However, technologies still primarily affect the productivity coefficients of the firms' production function, and are thus not necessarily related to specific technology paths such as those considered in process-based IAMs.
It is thus unclear whether these models are overly optimistic or pessimistic, as noted in the critiques of \citet{KreyEtAl2019LookingHoodComparison}.

Finally, in terms of the financial system, both the DSK-FIN and the ABMIAM incorporate a detailed representation of imperfect capital markets that may lead to bank failures. In particular, the DSK-FIN extension is focused around studying the financial fragility emanating from increasing climate damages. The ABMIAM also offers a heterogeneous banking sector, which also includes interbank loans, thus creating an impcit interbank network that may be subject to cascading crises and the types of systemic risk that is not captured in more common IAMs \citet{Monasterolo2020ClimateChangeFinancial}.

\section{Energy and Resource Modules}%
\label{sec:abiamenergy}
In this section, I address how the four considered models treat the energy-resource-environment nexus. The representations range from very stylized (GRSW and DSK) to more sophisticated (CFHS) in terms of the variety of energy sources used in electricity production.

Beginning with the raw sources of materials and primary energy sources, the DSK and ABMIAM models consider stylized primary energy types, involving an extracted fossil fuel (gas and coal in ABMIAM) and a form of renewable energy that functions without the requirement for fuel inputs.
In both these models, the extraction and provision of these fuels are outside of the model's boundary, and are thus infinitely available with exogenously given price processes.
While the GRSW model does not have an energy sector and primary energy sources, it does have a mining sector that extracts material resources in the southern region and sells them to the capital goods sector in the northern, yet here too there are no dynamics of mine-depletion as the purpose is primarily to study unequal exchange between north and south in the presence of local and global pollution.

CFHS goes beyond this, by considering not only the conversion of primary energy to electricity, but also the extraction of fossil fuels.
In particular, they consider seven distinct fuel types (coal, gas, oil, nuclear, hydro, wind and solar), with a regional fuel-extraction sector.
The presence of a fuel extraction sector is important, as the dynamics of reserve depletion through extraction and the shifts in regional extraction will impact the agents decisions about which type of power plant to invest in, what degree of emissions are possible, and how technological innovation interacts with depletion \citep[see debates in][]{HookTang2013DepletionFossilFuels, Capellan-PerezEtAl2014FossilFuelDepletion}.
In the CFHS model, each region has an empirically determined reserve of all fossil fuels.
The marginal cost of their extraction is increasing in the cumulative amount of extraction (modelled by a Rogner curve as in \citet{NordhausBoyer2003WarmingWorldEconomic}), reflecting the principle that the easiest-to-extract reserves are captured first.
Likewise, there is thus a regional shift as local reserves become depleted, favoring those regions with more abundant resources.
This is an important driver in evolving regional inequalities, in the sense of the unequal exchange studied in the GRSW model (though they do not include depletion dynamics), as investments and employment will increase in those regions with the cheapest to extract resources.

Turning to the transformation of primary energy into energy carriers, the DSK, ABMIAM and CFHS models all consider only electricity as a final energy carrier.
Notably, the CFHS model has a demand for fuel by households, but considers this to be part of the primary energy sector (there is no transformation to liquid fuels).
All three of these models consider heterogeneous power plants in terms of their fuel type, cost structure, energy and emission intensity.
For the DSK and ABMIAM the production is centrally cleared in a given period: sectors present their electricity demand and power plants are activated until this demand is met.
The price is then determined as a markup on the operating cost of the final power plant to be activated such that supply meets demand.
The CFHS model takes this one step further by considering the variability in renewable electricity generation.
They split each simulation step into a series of substeps with alternating zero and peak production of solar.
Markets are then cleared in the same procedure as the DSK and ABMIAM but on a subperiod base, which leads to an average price per simulation step.
Furthermore, fossil and nuclear plants are also considered stochastic, activating with a beta distribution proportional to their capacity factors in order to simulate the effects of maintenance which can lead to price jumps.\footnote{Such as simultaneous refueling and maintenance of nuclear plants in France in 2022 that increased the electricity price in the context of the gas crisis}

Across all models, investment in the replacement of power plants once they become obsolete, and the decision about the type of power plant to construct are similarly based on the net present value of each powerplant option, the candidate plant with the highest expected value (lowest lifetime cost) is then selected.
Where the models differ is in how innovation in the energy-sector occurs.
The DSK applies an endogenous innovation framework, which allows for reductions in the fixed costs of renewable power plants while for fossil-fuel plants there is an improvement in emission intensities.
By contrast, the CFHS considers an exogenous noisy decrease in the investment cost for solar and wind, to a specified minimum, as well as an increase in the substitutability of electricity for fuels in consumer demands.
While the technology process of the DSK is endogenous, it is unclear whether the reviewed technological innovations are overly optimistic or pessimistic, and thus whether they adequately respond to similar critiques of process-based IAMs \citep{KeppoEtAl2021ExploringPossibilitySpace}.

In summary, the energy sector remains rather stylized in three out of the four models, thus limiting the description of energy source substitution and technological advances in this respect, as well as the dynamics of fuel extraction and the multi-scale nature of energy flows through the economy \citep{GiampietroEtAl2012MetabolicPatternSocieties}.
However, the CFHS model showed that a more detailed representation of the energy sector is feasible within ABIAM, including an implementation of resource extraction in a multi-regional context.
Agent-based approaches have already been applied to energy-sector specific elsewhere, suggesting that more detailed energy-sector representations in ABIAM should be feasible \citep[see][for more information]{CastroEtAl2020ReviewAgentbasedModeling}.

\section{Climate Modules}%
\label{sec:abiamclimate}
This section addresses how ABIAMs model the environment and crucially the feedback mechanisms between socioeconomic system and environment. Table \ref{tab:climate} summarizes the climate and environmental modules of the four reviewed ABIAMs.
The climate modules of the four reviewed models use similar simplified models of the carbon cycles as in the IAMs based on \citet{Nordhaus2017RevisitingSocialCost}.
These climate boxes have been critiqued as over-simplifying the carbon cycle, in particular with relation to irreversible climate tipping points that lock-in certain temperature changes \citet{DietzEtAl2021AreEconomistsGetting}. 

Where the ABIAMs differentiate themselves is in modeling the damage feedback from increasing temperature anomalies, and in the case of the GRSW model also the effects of localised pollution.
As a reference point, consider the case of \citet{Nordhaus1992OptimalTransitionPath}, where a deterministic (quadratic) function of temperature gives a multiplier that is applied to the total production. For example, a given temperature might map to a 5\% reduction in total GDP compared to an ``absence-of-damages'' world.
This approach has been quite heavily critiqued as underestimating damage impacts and their spatial heterogeneity, and ignoring non-linear tipping points \citep{LentonEtAl2019ClimateTippingPoints, SteffenEtAl2018TrajectoriesEarthSystem}.
In contrast, each of the ABIAMs has a specific treatment of the size of the damage as well as its distribution amongst the model agents.
With the exception of the ABMIAM model, there are also multiple types of damage, generally including damage to capital stocks and labor reductions (the ABMIAM considers only shocks to the consumers' budgets) that are applied at the agent-level. 
This approach alone already differs from the cost-benefit IAMs.

Turning to the calculation of damage magnitude and distribution, there is significant heterogeneity between the models dependent on their intended purpose.
The DSK model directly aims to address the issue of the increasing variability and risk of rare climate events omitted in deterministic damage functions.
In this spirit, the increase in temperature anomaly changes the parameters of a beta distribution, leading to a higher mean damage and a stronger skew.
With each firm hit by a random draw affecting capital and labor productivity, \citet{LampertiEtAl2019AgentbasedIntegratedAssessment} conclude that ``under the same 'business-as-usual' emission scenario, roughly adherent to the Representative Concentration Pathway 8.5, the average climate shock from DSK and the damage function in DICE2013r \citep{Nordhaus2014EstimatesSocialCost} are vaguely similar: the 2100-$\mu$-level shock averages 5.4\% while the DICE damage function implies a GDP loss of approximately 5.2\%. However, aggregate impacts are radically different, with end-of-century projected output being around 90\% of the 'without-climate-change' scenario in DICE while amounting to 15\% in DSK (with shocks assumed to target labor productivity)''.

The choices made across models differ.
The DSK-FIN uses the deterministic damage function from \citet{Nordhaus2017RevisitingSocialCost}, applying the damage perturbed with noise to all firms, in order to facilitate comparison to the cost-benefit IAM research.
Meanwhile, the ABMIAM and the CFHS model aim to understand how heterogeneity in agents and shocks can lead to different results than when using the representative agent and aggregate damages.
In particular, the ABMIAM looks at the social cost of carbon and its associated optimal tax rate when shocks are distributed amongst individual agents in relation to their wealth, finding significant impacts of wealth and rent-inequalities on the social cost of carbon, such as increased poverty rates, leading to a deterioration of the economy.
CFHS aim to verify whether the aggregate damage functions arise when considering different agents and empirically calibrated deterministic micro-shocks.
To do so they take literature estimates of agricultural and labor shocks, and empirically estimate the damage of natural disasters for each of their 10 sub-regions. 
Running the model, they then study the shape of the aggregate damage function to compare it to cost-benefit type deterministic functions. 
They find that depending on remaining fuel availability and the speed of renewable energy growth, the aggregate damage function can take different shapes, and have differing magnitudes.
Finally, the GRSW model proposes that damages are proportional to the total stock of capital, while the frequency of shocks per period increases with the temperature anomaly.
Thus increasing the probability that one firm may be hit multiple times per period in a non-linear way. 
This is in conjunction with being the only model to consider localised pollution as well.
In the mining sector, workers' health, proxied by their labor productivity, decreases with the period of their employment in the mine, thus requiring ever more new labor.
They find that in an accelerated climate scenario, damages outstrip efforts to repair and innovate, leading to a long-term collapse of the economy. 

In summary, ABIAMs have begun addressing the underestimation of climate damages suggested by \citet{Stern2016EconomicsCurrentClimate}.
The benefit of the modular structure of ABMs is that many different forms and distributions of damage can be considered, such as regional damages, as applied in CFHS, or agent-specific damages as in the ABMIAM.
Thus models can easily incorporate on-going research on the micro-impacts of climate change \citep{CarletonHsiang2016SocialEconomicImpacts} and contribute to the ongoing debate on climate damage modelling \citep{DiazMoore2017QuantifyingEconomicRisks}.

\section{Policy Implications}%
\label{sec:abiampolicy}

The considered models differ markedly in the scope of implemented and tested policy options, depending on the respective authors' stated purpose.
With the exception of CFHS, each model considers some form of policy aimed at mitigating carbon emissions and hence the negative feedback effects from increases in $CO_2$ concentration.
Overarchingly, the benefit of the modular ABIAMs is that policies that have been suggested in the literature but are not typically possible in cost-benefit IAMs can be tested.
To highlight this, I consider the case of green financial regulation in the DSK-FIN model and the implementation of a global markets institution in the GRSW model.

The literature on the climate-finance nexus suggests that an active role of financial regulators can help shape climate related risk management \citep{Monasterolo2020ClimateChangeFinancial} and a low-carbon transition \citep{CampiglioEtAl2018ClimateChangeChallenges}.
The DSK-FIN model has a detailed set of heterogeneous banks that are subject to credit regulation. In this context, the authors test three regulatory policies: (i) green Basel requirements which excludes loans to green firms are excluded from banks capital requirements, (ii) green credit guarantees where the government guarantees the value of loans to green firms, and (iii) carbon-adjusted credit ratings where greener companies receive an improved credit rating. In the context of the model, the authors find that individually none of the policies manage the trade-off between climate mitigation and stable growth, with either a mitigation at the cost of more frequent banking crises, or a stable economy until climate damages become extreme enough to lead to collapse. 
In addition to the new financial regulations, the DSK model also includes standard leaning-against-the wind policies such as unemployment benefits and taxes on firm profits, as well as credit-multipliers to limit total debt.

In contrast to the financial system regulation, \citet{GerdesEtAl2022LaborEnvironmentGlobal} use the GRSW model to explore the implications of a ``civilized market institution'' that has the power to fine capital-goods companies for emissions and mines for local pollution, and redistribute the collected fines in the form or subsidies for retrofitting firms' capital to reduce their pollution and emission intensities. Their results suggest that fines alone, which may include items such as carbon taxes, are insufficient to counter natural disaster damage nor unequal exchange, regardless of the severity of the fines. On the other hand, when these fines are used to subsidize innovation activities in the form of retrofitting capital, does the value chain become sustainable over the long run. All cases are cost-neutral, as subsidies are funded by sanctions, thus converging to zero as emissions are mitigated. They find that the most promising results include a North-South transfer as capital firms in the north still invest in carbon reductions at the same rate, but significantly boosts mitigation in mines.
In a similar vein to the DSK, the model contains unemployment benefits and taxes on firm profits as well.

The ABMIAM of \citet{SafarzynskavandenBergh2022ABMIAMOptimalClimate} considers the more common policy of carbon taxation. Specifically, they use an adapted form of the analytically derived optimal policy by \citet{RezaiVanderPloeg2016IntergenerationalInequalityAversion}, with revenues being redistributed as equal lump-sum payments to citizens. They then analyze to what degree these taxes may reduce inequalities versus the same scenarios in their absence. Their findings suggest that income inequality can be addressed by these policies, while consumption and wealth inequality are not strongly affected.\footnote{Consumption inequality is addressed in an unjust economy where damages are allocated to the lowest wealth households, as this essentially functions as a non-damaged income such that households can maintain their consumption where otherwise they would be in poverty}

The policies highlighted in the DSK-FIN and GRSW model show the potential for ABIAM in policy analysis because ABIAM are able to also incorporate regulatory policies and a wide array of policy mixes. This would allow for consultation with policymakers to study in more expansive terms policy interactions and implications. It also allows to address some of the critiques that a focus on carbon pricing omits the interaction with innovation and diffusion processes \citep{RosenbloomEtAl2020WhyCarbonPricing}, as well as the interactions and trade-offs with other sustainable development goals such as equity \citep{GeelsEtAl2016BridgingAnalyticalApproaches}.
At the same time, \citet{KeppoEtAl2021ExploringPossibilitySpace} note that policies alternative to the commonplace carbon taxation or trading are also possible in process-based IAMs, and suggest that the focus is driven by how models are meant to be used and what policymakers have demanded.

\section{The Path Ahead}%
\label{sec:nextsteps}
While ABIAMs have made advances in understanding the micro-impacts of climate shocks, the effects of inequality on policy, and the interlinkages between the economy, the financial system and climate feedbacks, there is much yet to be done to match the detail of process-based IAMs.
In first place, the current set of ABIAMs should continue developing to incorporate features common to process-based IAMs, such as a fine-grained representation of the energy system, as started in CFHS, a multi-sectoral structure beyond stylized consumption and capital goods, and a connection with land-use and change models.
Several of these areas have already been addressed by agent-based models, such as electricity markets \citep[see][for reference]{CastroEtAl2020ReviewAgentbasedModeling}) or multi-sectoral and multi-regional structures as in the LAGOM models \citep{WolfEtAl2013MultiagentModelSeveral}, ACCLIMATE \citep{OttoEtAl2017ModelingLossPropagationGlobal} or \citet{CaianiEtAl2018EffectsFiscalTargets, CaianiEtAl2019EffectsAlternativeWage}.
Beyond the addition of detail and combination of these models, there are several areas worth highlighting. 

\citet{BardiPereira2022Limits50Years} point out that ``In recent years, global warming and climate change have become the main focus of the environmental movement. That may have led to the importance of resource depletion being neglected - more evidence of the importance of an integrated approach.''
A review by \citet{PollittEtAl2010ScopingStudyMacroeconomic} confirms these suspicions by reviewing 60 energy-environment-economics models and coming to the conclusion that ``consumption of material inputs is largely unexplored within a dynamic macroeconomic framework''.
More recent reviews by \citet{PauliukEtAl2017IndustrialEcologyIntegrated} find that IAMs have missing material-energy-economy linkages in their descriptions of installed capital and infrastructures.
Indeed, also in the reviewed ABIAMs only GRSW consider the extraction of a physical resource to construct capital, while also not treating its end-of-lifecycle implications.
Given the scale of resource extraction and the different required resources, even a saturation of installed capital levels would require large quantities of new virgin materials in developing countries, and for the energy transition \citep{WiedenhoferEtAl2019IntegratingMaterialStock, WiedenhoferEtAl2021ProspectsSaturationHumanity, WatariEtAl2019TotalMaterialRequirement}, which may make a full transition infeasible \citep{Michaux2021MiningMineralsLimits}.
Noting here that the extraction and processing of raw materials is typically an emission-intensive process.
Finally, also the end-of-life waste and recycling may become an important factor in a circular economy system with reduced virgin material extraction, this is also underappreciated in current models \citep{McCarthyEtAl2018MacroeconomicsCircularEconomy}.
The detailed and heterogeneous structure of ABMs should learn from and implement insights from industrial ecology to assess not just the socioeconomic but also the material dimensions of a climate transition.
In particular, I suggest considering modeling at the stock-flow-practices nexus \citep{HaberlEtAl2021StocksFlowsServices}, where the practices and demands of individual consumers become important in the requirements for various materials. Modeling individual agents also allows for interactions and social phenomena on the demand-side that may change the path of the climate transition \citep[e.g. see the review of][for some ABM-based approaches]{CastroEtAl2020ReviewAgentbasedModeling}, where there are unexplored policy avenues \citep[e.g. see][]{FitzpatrickEtAl2022ExploringDegrowthPolicy}.

All of the ABIAMs reviewed in this model show that there are distributional effects of climate change, especially when considering the differential impacts of climate feedbacks both regionally in the CFHS and GRSW models, and amongst individuals in ABMIAM and DSK.
More broadly, ABM-type models may be able to incorporate more indicators and analyses relating to the interaction amongst different Sustainable Development Goals such that policymakers and citizens may assess and decide on trade-offs.
Such analyses may include assessing the climatic impacts alternative forms of policy that address issues of inequality and may restructure the socio-economic sphere such as proposals for universal basic income \citep[see][for recent reviews]{HoynesRothstein2019UniversalBasicIncome, BanerjeeEtAl2019UniversalBasicIncome}. 

Finally, additions and enlargements of Agent-based Models can run into computational issues.
Thus one wishes to carefully choose where to model heterogeneity to capture key aspects of the overall system behavior \citep{KeppoEtAl2021ExploringPossibilitySpace}.
To facilitate this, a process of \textit{agentization} should be considered \citep{GuerreroAxtell2011UsingAgentizationExploring}.
This involves beginning the modeling exercise with a system dynamics approach that aims to capture overall system behavior at an aggregate level, and then slowly replacing sectors or aggregations with individual agents.
This would facilitate both the computational exercise, validation, and communicability of the results of agent-based models. 

\section{Conclusion}
In this paper, I have reviewed four distinct Agent-based Integrated Assessment models as examples of an alternative framework with which to develop economy-energy-environment models that address the global climate crisis. ABIAMs have several advantages to aggregated equilibrium-based models, and are able to respond to the critiques directed at IAMs. These advantages include: (1) heterogeneity within and across different groups of agents, which enables the study of distributional and equity issues. (2) The allocation of damages across micro-agents in the model with differentiated magnitudes, and in the case of DSK also increasing variability. (3) ABIAMs have an easier time representing a financial system populated by multiple banks. This allows for an explicit consideration of how financial regulation may impact the transition to a low-carbon economy through different investment channels, and the associated effects on financial system risk, which leads to (4) a wider range of policy-mixes can be examined, including their distributional impacts.  

While there are a multitude of benefits to the agent-based modeling approach when applied to integrated assessments \citep[see also]{BalintEtAl2017ComplexityEconomicsClimate, LampertiEtAl2019AgentbasedIntegratedAssessment}, there are many avenues along which these models must yet progress. These include more detailed descriptions of energy systems and the inclusion of land-use and production sectors to match current process-based IAMs. Furthermore, ABIAMs should consider the inclusion of insights from industrial ecology and the full material lifecycle of capital and infrastructure, as well as the testing of alternative policies and new policy mixes to address some of the open questions and critiques surrounding IAMs more generally.

\section{Acknowledgements}
This research was conducted within the Econophysics \& Complex Systems Research Chair, the latter under the aegis of the Fondation du Risque, the Fondation de l’Ecole polytechnique, the Ecole polytechnique and Capital Fund Management. I also acknowledge the support from the New Approaches to Economic Challenges Unit at the Organization for Economic Cooperation and Development (OECD).

\newpage
\appendix
\section{Intercomparison of ABIAMs}
\begin{center}
\footnotesize
\begin{xltabular}{\linewidth}{>{\raggedright\arraybackslash}X>{\raggedright\arraybackslash}X>{\raggedright\arraybackslash}X>{\raggedright\arraybackslash}X>{\raggedright\arraybackslash}X>{\raggedright\arraybackslash}X}
    \caption{Overview of the General Structure of four Agent-based Integrated Assessment Models. Details on the energy and resource sector are in Table \ref{tab:energyresources}, the climate module in Table \ref{tab:climate}, and policy institutions in Table \ref{tab:policy}}\label{tab:macro}\\
    \toprule
     & \textbf{DSK} & \textbf{DSK-FIN} & \textbf{ABMIAM} & \textbf{CFHS} & \textbf{GRSW} \\
     \midrule
     \endfirsthead
     & \textbf{DSK} & \textbf{DSK-FIN} & \textbf{ABMIAM} & \textbf{CFHS} & \textbf{GRSW} \\
     \midrule
     \endhead
     \midrule
     \mc{6}{r}{\textit{Table continues on next page}}\\
    \endfoot
    \endlastfoot
    Key References & \citet{LampertiEtAl2018FarawayCloseCoupled, LampertiEtAl2020ClimateChangeGreen} & \citet{LampertiEtAl2019PublicCostsClimateinduced, LampertiEtAl2021ThreeGreenFinancial} & \citet{SafarzynskavandenBergh2022ABMIAMOptimalClimate} & \citet{CzuprynaEtAl2020AgentBasedApproachIntegrated} & \citet{GerdesEtAl2022LaborEnvironmentGlobal} \\
    Macroeconomic Framework & K+S Model & \citet{LampertiEtAl2018FarawayCloseCoupled} &  \citet{SafarzynskavandenBergh2017FinancialStabilityRisk, SafarzynskavandenBergh2017IntegratedCrisisenergyPolicy}& No prior framework & Influenced by \citet{RengsScholz-Wackerle2019ConsumptionClassEvolutionary}\\
    \midrule\mc{6}{l}{\textbf{General Properties}}\\\midrule
    Stock-flow Consistent & Yes & Yes & Unspecified & Unspecified & Unspecified \\
    Timestep & Quarter & Quarter & Year & Year & Monthly with annual-only events\\
    Regions & Single & Single & Single & 10 regions: Africa, Japan, China, India, Rest of Asia, Europe, North America, Central and South America, Commonwealth of Independent States, Middle East & Two: Global North (with capital goods), Global South (with mines)\\

    \midrule\mc{6}{l}{\textbf{Consumption}}\\\midrule
    Reference & \citet{LampertiEtAl2018FarawayCloseCoupled} &\citet{LampertiEtAl2018FarawayCloseCoupled} & \citet{SafarzynskavandenBergh2017IntegratedCrisisenergyPolicy}& \citet{CzuprynaEtAl2020AgentBasedApproachIntegrated}& \citet{GerdesEtAl2022LaborEnvironmentGlobal}\\
    Goods & Single C-Good & Single C-Good & Multiple goods. Each firm offers good with differentiated quality depending on maximum attainable quality and duration of producing a given good & Multiple goods: agriculture, textiles, chemicals, other manufacturing, transport, and other services & Single C-Good\\
    Production & Constant returns to scale with labour and capital vintages &Constant returns to scale with labour and capital vintages & CES production function with labour, capital and energy & CES with capital, labour and energy & Leontief function with capital and labour \\
    Factor Demand & Proportional to target production & Proportional to target production & Cost minimization of CES given production target & Cost minimization of CES given production target & Proportional to target production\\
    Supplied Quantity & Expected demand with adjustment for inventory&Expected demand with adjustment for inventory & Weighted average of current sales and actual demand & Tatonnement: increase if excess demand and price above market  \citep{AssenzaEtAl2015EmergentDynamicsMacroeconomic} & Target production depends on expected demand with adjustment for target inventory \\
    Pricing & Variable markup based on market share &Variable markup based on market share & Variable markup on cost based on past market power. Costs include variable cost and fixed cost of producing quality good & Tatonnement: increase if below market and excess demand, decrease if reverse \citep{AssenzaEtAl2015EmergentDynamicsMacroeconomic} & Tatonnement: Price is adjusted downward if strongly lower sales, and upward if strong excess demand. Bounded from below by production costs. Adoption of a new price is successful with a fixed probability \\
    Investment & Based on expected demand, with scrapping based on expected gains. & Based on expected demand, with scrapping based on expected gains. & Based on expected demand & Based on excess of capital factor demand over existing capital& Target based on expected demand.\\
    Household Demand & All income consumed & All income consumed & Consume based on target wealth-to-permanent-income ratio& Part of income (fixed propensity) plus any savings & Budget is wage income plus proportion of savings\\
    Household Choice & N/A & N/A & Intensity of choice function with utility per product. Product-utility is a Cobb-Douglas type aggregation of quality, price and number of other buyers & Stone-Geary utility function over $n$ goods, with a required minimum consumption. Energy consumption is split into electricity and fuels based on CES, then fuels is split into coal, gas and oil based on CES & Household has shortlist of preferred firms, with 25\% change household might change firm to a lower price one. With 25\% chance household replaces firm that didn't satisfy demand with random alternative\\
    Market Protocol & Not explicit & Not explicit & Not explicit & Regional clearing except transport and other services. Consumers visit producers (unspecified order) adjusting demands based on prices as clearing proceeds& Two purchasing rounds: a fraction of desired demand bought in round one, the remainder in round two. If a firm cannot satisfy demand, a household tries another firm until satisfied or no supply left.\\

    \midrule\mc{5}{l}{\textbf{Capital}}\\\midrule
    Reference & \citet{LampertiEtAl2018FarawayCloseCoupled} & \citet{LampertiEtAl2018FarawayCloseCoupled} & \citet{SafarzynskavandenBergh2017IntegratedCrisisenergyPolicy}& \citet{CzuprynaEtAl2020AgentBasedApproachIntegrated} & \citet{GerdesEtAl2022LaborEnvironmentGlobal}\\
    Capital Goods & Multiple vintages: labour-productivity, energy-efficiency, emission intensity &Multiple vintages: labour-productivity, energy-efficiency, emission intensity & N/A no explicit sector & Single type of capital & Single type of machine\\
    Production Technology & Constant returns to scale with labour and energy & Constant returns to scale with labour and energy & N/A & CES with capital, labour and energy & Leontief with capital, labour and resources \\
    Technological Change & \citet{NelsonWinter1982EvolutionaryTheoryEconomic} imitation-innovation process based on R\&D investment \citep[See ][]{DosiEtAl2010SchumpeterMeetingKeynes} & \citet{NelsonWinter1982EvolutionaryTheoryEconomic} imitation-innovation process based on R\&D investment \citep[See ][]{DosiEtAl2010SchumpeterMeetingKeynes} & \citet{NelsonWinter1982EvolutionaryTheoryEconomic} two-step process for improving CES technical coefficients. Maximum product quality and technical coefficients increase exogenously over time & Exogenous noisy growth in CES technical coefficients & None. Fixed technical production coefficients \\
    Prices & fixed markup & fixed markup& N/A since not a sector& Tatonnement: increase if below market average and excess demand \citep{AssenzaEtAl2015EmergentDynamicsMacroeconomic} &  Tatonnement: Price is adjusted downward if strongly lower sales, and upward if strong excess demand. Bounded from below by production costs. Adoption of a new price is successful with a fixed probability \\
    Quantity & Proportional to expected demand & Proportional to expected demand & N/A & Based on expected demand, subtracting own requirements first & Target depends on expected demand with an inventory adjustment\\
    Market Protocol & Choice by price, productivity, energy-efficiency. Can consume froma subset of producers. Delivery at end of period & Choice by price, productivity, energy-efficiency. Can consume froma subset of producers. Delivery at end of period & N/A since not a sector & \citet{AssenzaEtAl2015EmergentDynamicsMacroeconomic}: each consumer connected to fixed number of producers, purchasing from lowest producers in order of lowest price first. Excess demand is partially satisfied. Once fixed network is completed, additional demand is fulfilled by excess supply or proportionally allocated. Adjustment for import/export: if a region has an export surplus, imports are preferred & Consumption good firms have shortlist of capital firms, updated every 3 months with 25\% change household might change firm to a lower price one. With 25\% chance household replaces firm that didn't satisfy demand with random alternative. Firm attempts to buy equal fraction of demand from all sellers, if unsatified continues until at least 95\% satisfied or no more stock\\

    \midrule\mc{6}{l}{\textbf{Labour \& Household Income}}\\\midrule
    Reference & \citet{LampertiEtAl2018FarawayCloseCoupled} & \citet{LampertiEtAl2018FarawayCloseCoupled} & \citet{SafarzynskavandenBergh2022ABMIAMOptimalClimate}&\citet{CzuprynaEtAl2020AgentBasedApproachIntegrated}& \citet{GerdesEtAl2022LaborEnvironmentGlobal}\\
    Supply & Fixed & Fixed & Fixed & Fixed Regionally & Fixed Regionally\\
    Differentiation & None & None & None & None & Global south adjusts labour productivity based on local pollution and duration of employment\\
    Wages & Single market wage based on productivity, price, unemployment & Single market wage based on productivity, price, unemployment & Wages are fixed, with a proportional increase when demand exceeds supply of labour& Unspecified & All firms adjust wage based on change in price of their good over prior 12 months, with a maximum adjustment rate. Downward rigid. Mining employee wages are adjusted for productivity\\
    Other income & None & None & A share of energy of energy and capital rents. Interest on deposits. & Dividends from shareholdings in companies & half of firm profits are distributed regionally in proportion to their existing savings. Excess R\&D funds are distributed analogously\\
    Market Protocol & Not explicit & Not explicit & Randomly allocated workers  & Randomly allocated within regions & Regional markets with random matching. Capital and consumption goods have a maximum amount of new hires, and maximum proportion of labour that can be fired. Mines\\

    \midrule\mc{6}{l}{\textbf{Credit \& Financing}}\\\midrule
    Reference & \citet{LampertiEtAl2018FarawayCloseCoupled} & \citet{LampertiEtAl2021ThreeGreenFinancial} based on \citet{DosiEtAl2015FiscalMonetaryPolicies} & \citet{SafarzynskavandenBergh2017IntegratedCrisisenergyPolicy}&\citet{CzuprynaEtAl2020AgentBasedApproachIntegrated} & \citet{GerdesEtAl2022LaborEnvironmentGlobal}\\
    Credit Demand & Desired investment net of cash & Desired investment net of cash & Desired investment net of cash & Desired investment & Desired investment net of cash\\
    External Finance & Bank loans & Bank loans, and government bonds for government purchased by banks & Bank loans & Household investment & Bank loans\\
    Firm Bankruptcy & Negative liquid assets or zero market share. Replaced by firm's representing industry averages & Negative liquid assets or zero market share & Inability to pay back loans, high market share firms can extend loan payback by some periods. Capital goes to the bank, which resells to new entrants with fixed probability & Consumer \& capital firms go bankrupt if there is a lack of demand or lack of production factors. Fuel extraction companies go bankrupt if all resources are exhaused. Ownership covers the losses & No production for 12 months\\
    Firm Entry & Replace bankrupt firm by industry average firm & Replace bankrupt firm by industry average & new firm enters with fixed probability, offering random product quality with production technology greater than best current technology. Firm demands start-up loan for initial investments, successfully granted with a fixed probability. First period has a zero markup price & Unspecified & Unspecified\\
    Credit Supply & Pecking-order based on net-worth-to-sales. Upper bound to credit based on debt-to-sales& Bounded by each banks' equity, which is subject to a Basel-II capital rule. Pecking order based on firm's credit worthiness&Firm receives loan if below exogenous debt-to-equity ratio. & Household decide on overall planned investment (income net of consumption) & Unbounded \\
    Interest rate & Markup on central bank rate & Risk premium based on client position in credit ranking (markup is based on quartile within the banks clientele)& The same offer from all banks, with electricity having a lower interest rate & No interest. Firms pay out profits & Fixed rate\\
    Regulation & Maximum credit set by credit multiplier rule & Time-varying capital adequacy ratio &Minimum level of reserves held at central bank, with fixed minimum and higher reserves depending on deposits & None & None\\
    Bank Bankruptcy & None & Equity (net worth) is negative& Equity or reserves are negative. Lending banks write off the loans to defaulted banks & None & None\\
    Bank Entry & None & Bankrupt bank is bailed out by the government. Bailout up to a (fraction of the equity of the smallest incumbent) & None & None & None\\ 
    Market Protocol & Single bank, pecking-order credit & Each bank has a pecking order process for its list of clients & Loans granted based on debt-to-equity ratio. If a bank has insufficient funds to grant a loan, it asks other banks for loans (order of liquidity) until liquidity requirements are satisfied & Households distribute planned investment among companies based on their current ownership shares and value of companies planned capital increase. If planned investment exceeds demand, money is distributed back to owners & All credit granted if profit rate is larger than interest rate adjusted for credit-lenience\\
    \bottomrule
\end{xltabular}
\end{center}

\begin{center}
\footnotesize
\begin{xltabular}{\linewidth}{>{\raggedright\arraybackslash}X>{\raggedright\arraybackslash}X>{\raggedright\arraybackslash}X>{\raggedright\arraybackslash}X>{\raggedright\arraybackslash}X}
\caption{Overview of the Energy and Resource Module of four Agent-based Integrated Assessment Models}\label{tab:energyresources}\\
    \toprule
     & \textbf{DSK \& DSK-FIN} & \textbf{ABMIAM} & \textbf{CFHS} & \textbf{GRSW} \\
     \cmidrule{1-5}
     \endfirsthead
     & \textbf{DSK \& DSK-FIN} & \textbf{ABMIAM} & \textbf{CFHS} & \textbf{GRSW} \\
     \midrule
     \endhead
     \midrule
     \mc{5}{r}{\textit{Table continues on next page}}\\
    \endfoot
    \endlastfoot
    Reference & \citet{LampertiEtAl2018FarawayCloseCoupled} & \citet{SafarzynskavandenBergh2022ABMIAMOptimalClimate} based on \citet{SafarzynskavandenBergh2011IndustryEvolutionRational, Safarzynska2012ModelingReboundEffect} & \citet{CzuprynaEtAl2020AgentBasedApproachIntegrated}&\citet{GerdesEtAl2022LaborEnvironmentGlobal}\\
    \midrule
    \mc{5}{l}{\textbf{Energy Markets}}\\
    \midrule
    Primary Energy Types & representative fossil fuel, representative renewable flow & three fuels: (1) increasing unit cost over time, (2) decreasing unit costs over time by Brownian motion, (3) constant price. Associated with gas, coal and renewable energy & Coal, gas, oil, nuclear, hydro, wind and solar&N/A\\
    Producer Heterogeneity & green (renewable) and brown (fossil) power plants with capital vintage dependent cost-structure, thermal efficiency, environmental impact. Production requires only capital, and fossil fuel for brown power plants (one unit per unit electricity) & heterogeneous plants differentiated by: age, productivity, energy source, installed capacity, maximum lifespan, capacity factor & Heterogeneous plants: fuel type, production capacity, storage capacity (solar and wind), lifetime, capacity factor, operation cost factor, electricity transmission loss factor, thermal efficiency factor & N/A\\
    Production Technology & Production requires only capital, and fossil fuel for brown power plants (one unit per unit electricity) & Cobb-Douglas function of capital, labour and fuel with plant-specific TFP inversely related to thermal efficiency increased by Gaussian every period and fixed exponents (substitution factors). & Fixed capacities. Planned electricity fixed for nuclear, beta distribution for all other types (maintenance and weather) with a shape proportional to capacity factors. Solar has double capacity in daytime, and zero at night. Maximum supply augmented by storage and reserve rates for combustion plants&N/A\\
    Production Capacity & Unitary per plant & Plant-specific chosen at creation of plant & Plant-specific&N/A\\
    Production Cost & price of fossil fuels in relation to thermal efficiency for brown plants, zero for green plants & cost of labour, fixed operating costs, fuel costs & operation cost proportional to production, plus labour cost&N/A\\
    Pricing & fixed markup on the marginal producers cost (supplier of last unit) & Inverse demand function with markup & Market clearing in each sub-period, with stress factors if demand > maximal provision&N/A\\
    Energy Storage & No & No & Yes, for solar and wind plants&N/A\\
    Final Energy Types & Electricity & Electricity & Electricity (Fuel considered a direct primary purchase)&N/A\\
    Market Structure & Central authority activating plants until demand is met & Cournot game: each plant produces to maximize profits given & Time split into sub-steps (stochastic generation profiles) & N/A \\
    Depreciation & All plants have a fixed lifetime & Plants have a fixed lifetime & Plants have a fixed lifetime, each period oldest part of capacity for a plant is depreciated.&N/A\\
    Physical Investment & Cost of new brown plant is zero, while new green plant has vintage-dependent fixed cost. Expansion is done when demand exceeds maximum production. Green plants are preferred as long as their fixed cost is less than present value of the cost of the most efficient brown plant & Once a fixed lifetime is reached, a plant exists and the owner invests in a new plant. The type of fuel is chosen based on expected profits from that fuel. A new plant receives a loan to construct, this has to be paid back at the end of its lifetime & Each period a share of old capacity is replaced by investment if maximum expected demand cannot be met post-depreciation. Maximal capacity increases are bounded from above. Solar and wind also invest in storage if electricity prices are volatile.&N/A\\
    Technological Innovation & Fraction of toal past sales invested in R\&D, allocated in proportion to revenue generated. \citet{NelsonWinter1982EvolutionaryTheoryEconomic} two-step process: (1) draw for successful innovation, (2) beta-distribution draw or proportional improvement & None - fixed Cobb-Douglas exponents & Overnight investment costs for solar and wind decrease to a specified floor. Substitution of electricity for fuels inreases to match intra-fuel substitutability&N/A\\
    Final Demand & C-Good and K-good sectors & C-good market & All sectors except electricity producers&N/A\\
    Market Protocol & Plants activated in order of cost (green first) until demand is met & Plants choose production to maximize profit based on Cournot game with linear inverse demand function (coefficient adapte to guarantee i). Factor demands are based on marginal productivity & In each region, time is in sub-steps with half as night (no solar, low demand) and half as day (higher demand) and cleared each time to develop an average price for the period.&N/A\\
    \midrule
    \mc{5}{l}{\textbf{Resource Markets}}\\
    \midrule
    Types of Resources & Representative fossil fuel & Three stylized fuels & Coal, Crude Oil, Natural Gas & Representative extracted resource\\
    Stock and Extraction & Infinite stock & Infinite stock & Finite, regional stocks. Marginal costs of extraction follow Rogner curve. Planned production is adjusted for regional depletion rates (easy to extract first). & Infinite stock\\
    Production Technology & N/A & N/A & Capital gives capacity, fixed total labour. Capital adjusted based on expected vs. realized demand. No innovation. & Leontief dependent on labour and capital\\
    Price & Exogenous variable & Geometric Brownian Motions & Intersection of supply and vertical demand (short-term inelasticity) &   Tatonnement: Price is adjusted downward if strongly lower sales, and upward if strong excess demand. Bounded from below by production costs. Adoption of a new price is successful with a fixed probability \\
    Market Protocol & N/A & N/A & Centrally cleared per fuel (coal, crude, gas). Supplied at the marginal extraction cost, consumers bid quantities and acecpt market price. If demand>planned supply, the highest price is taken and excess demand distributed over suppliers (assuming below maximum extraction). If demand>max.extraction, proportional rationing. &  Capital good firms have shortlist of capital firms, updated every 3 months with 25\% change household might change firm to a lower price one. With 25\% chance household replaces firm that didn't satisfy demand with random alternative. Firm attempts to buy equal fraction of demand from all sellers, if unsatified continues until at least 95\% satisfied or no more stock\\
    \bottomrule
\end{xltabular}
\end{center}

\begin{center}
\footnotesize
\begin{xltabular}{\linewidth}{>{\raggedright\arraybackslash}X>{\raggedright\arraybackslash}X>{\raggedright\arraybackslash}X>{\raggedright\arraybackslash}X>{\raggedright\arraybackslash}X>{\raggedright\arraybackslash}X}
\caption{Overview of the Implementation of Climate Modules in four Agent-based Integrated Assessment Models}
\label{tab:climate}\\
\toprule
& \textbf{DSK} & \textbf{DSK-FIN} & \textbf{ABMIAM} & \textbf{CFHS} & \textbf{GRSW} \\
     \cmidrule{1-6}
     \endfirsthead
& \textbf{DSK} & \textbf{DSK-FIN} & \textbf{ABMIAM} & \textbf{CFHS} & \textbf{GRSW} \\
\midrule
     \endhead
     \midrule
     \mc{6}{r}{\textit{Table continues on next page}}\\
    \endfoot
    \endlastfoot
    Reference Paper & \citet{LampertiEtAl2020ClimateChangeGreen} &\citet{LampertiEtAl2019PublicCostsClimateinduced}&\citet{SafarzynskavandenBergh2022ABMIAMOptimalClimate}&\citet{CzuprynaEtAl2020AgentBasedApproachIntegrated}&\citet{GerdesEtAl2022LaborEnvironmentGlobal}\\
     \midrule\mc{6}{l}{\textbf{Climate System}}\\\midrule
    Reference Framework & C-ROADS model of \citet{StermanEtAl2012ClimateInteractiveCROADS, StermanEtAl2013ManagementFlightSimulators}. Similar to \citet{Nordhaus1992OptimalTransitionPath}& Single-equation framework similar to \citet{MatthewsEtAl2009ProportionalityGlobalWarming, MatthewsEtAl2012CumulativeCarbonPolicy}&\citet{Nordhaus2017RevisitingSocialCost}&\citet{Petschel-HeldEtAl1999TolerableWindowsApproach} & Single equation framework\\
    Timescale & Annual & Quarterly (like model) & Annual & Annual & Monthly\\
    Pollution Sources & Consumption Goods Sector, Capital Goods Sector, and Energy Sector emit CO2 & Consumption Goods Sector, Capital Goods Sector, and Energy Sector emit CO2& Energy sector (fuel-specific) emits CO2& Energy sector (fuel-specific). \red{Unclear} Emits CO2 & Capital firms in the global North emit CO2. Mines in the global South cause local Pollution\\
    Description & Two-layer model with two loops: (1) increased ``natural'' primary CO2 production with CO2 levels, (2) oceans' capacity to uptake carbon falls with CO2 concentration increases. Radiative forcing determines mean temperature increase.& Fixed ratio for change in temperature following change in cumulative emissions& Changes in atmospheric carbon depend on existing stock and past emissions with a permanent and transient part. Stock of carbon affects global mean temperature anomaly\newline $T=\omega\frac{\ln \frac{E_t}{E_{pre}}}{\ln 2}$&Three-equation carbon cycle. Cumulative emissions increase by economy, leading to increase in carbon concentration (cumulative + annual emission - difference to prehistoric values). Temperature change is proportional to carbon intensity vs. prehistoric and temperature vs. prehistoric. Regional temperatures differ, but increase by the same as global mean temperature & Emission increases monthly, and decreases as a fixed proportion of initial concentration. Local pollution of the global south depends on mine production rate and their pollution coefficients\\
    
    \midrule\mc{6}{l}{\textbf{Damage Feedback}}\\\midrule
    Types of Damage & Capital stocks, labour productivity, inventories, energy-efficiency & Labor productivity, capital stocks& Consumer budgets & Agricultural, Labour, Natural Disasters & Capital stocks, Labor productivity of workers in mines due to local pollution\\
    Damage Distribution & Beta distribution of damage proportion (uni-modal right-skewed), with a mean increasing in temperature, and the right tail increasing with temperature variability (higher risk of extreme events) &  Damage follows \citet{Nordhaus2017RevisitingSocialCost} deterministic damage function: a quadratic function of temperature levels with two response parameters & Total damage proportional to increase in temperature\newline $\left( 1 + \zeta_1T_t^{\zeta_2} \right)^{-1}$& Deterministic functions. Agriculture: literature estimates for Europe/Africa/Rest-of-World. Labour: quadratic function of temperature difference to $13^\circ C$. Natural disasters: regional regression estimates of USD bn. loss per degree increase in temperature & Damage magnitude is a proportion of total capital level, monthly number of disasters (hits) is a nonlinear function of current emissions compared to the baseline, parameterized by an acceleration rate. \\
    Damage Allocation & Each firm is hit with a random draw affecting capital or labour productivity & All firms are hit by the aggregate shock modified by a small Gaussian noise parameter & Based on modification of \citet{DennigEtAl2015InequalityClimateImpacts}: consumption post damage is reduced by consumer-specific share of total damage, dependent on the share of wealth (through wealth-elasticity of damage)&Agricultural damages are reductions in all CES coefficients for Agriculture sector. Labour damages are reductions to labour efficiency of capital/consumer good companies. Natural disasters: reduce all firms' capital, reduce output of consumer and capital good firms & Each hit hits a random firm (capital, consumption or mine) allowing for multiple hits per month. Firms lose capital equivalent to the damage value. Local pollution per mine decreases the mine's employees productivity (from 1 down to a minimum of 1/3). Workers are replaced every 30 years (same employer and savings)\\
    Tipping Points & Implicit due to higher probability of extreme events & None & None & Non-linearity in data, but no tipping points & None\\
    \bottomrule
\end{xltabular}
\end{center}

\begin{center}
\footnotesize
\begin{xltabular}{\linewidth}{>{\raggedright\arraybackslash}X>{\raggedright\arraybackslash}X>{\raggedright\arraybackslash}X>{\raggedright\arraybackslash}X>{\raggedright\arraybackslash}X>{\raggedright\arraybackslash}X}
\caption{Overview of the Policy Experiments of four Agent-based Integrated Assessment Models}\label{tab:policy}\\
\toprule
 & \textbf{DSK} & \textbf{DSK-FIN} & \textbf{ABMIAM} & \textbf{CFHS} & \textbf{GRSW} \\
\cmidrule{1-6}
\endfirsthead
 & \textbf{DSK} & \textbf{DSK-FIN} & \textbf{ABMIAM} & \textbf{CFHS} & \textbf{GRSW} \\
 \midrule
 \endhead
 \midrule
\mc{6}{r}{\textit{Table continues on next page}}\\
\endfoot
\endlastfoot
    Reference & \citet{LampertiEtAl2018FarawayCloseCoupled, LampertiEtAl2020ClimateChangeGreen} & \citet{LampertiEtAl2019PublicCostsClimateinduced, LampertiEtAl2021ThreeGreenFinancial} & \citet{SafarzynskavandenBergh2022ABMIAMOptimalClimate} & \citet{CzuprynaEtAl2020AgentBasedApproachIntegrated} & \citet{GerdesEtAl2022LaborEnvironmentGlobal} \\
\midrule\mc{6}{l}{\textbf{Government}}\\\midrule
    Fiscal Policy (Outflow)& Unemployment benefits as a fixed fraction of current market wage & Unemployment benefits as a fixed fraction of market wage. Bank bailout costs & None & None & Unemployment benefits as 80\% of mean regional wage\\
Budget Deficits & Taxes on firm profits & Taxes on firm incomes and worker incomes. Government pays interest on its bonds to banks and the central bank, which are issued if deficits are positive. & None & None & Taxes on firm profits\\
\midrule\mc{6}{l}{\textbf{Central Bank}}\\\midrule
    Interest rate & fixed baseline rate & Taylor type rule with target inflation and unemployment rate& Fixed rates & None & Fixed baseline rate\\
    Credit regulation & sets credit multiplier (on deposits) to limit total debt & Bank specific credit multiplier depending on risk weighted assets and bank equity & Minimum reserves and fraction of deposits. Critical Debt-to-equity value for loan-granting & None & None \\
\midrule\mc{6}{l}{\textbf{Policy Experiments}}\\\midrule
    Fiscal & No Experiments & No experiments & Carbon Tax determined from the social cost of carbon in DICE models, the formula proposed by \citet{RezaiVanderPloeg2016IntergenerationalInequalityAversion}. Revenues are distributed as equal lump-sum to citizens & No Experiments & Global civilized market institution: sets base fine for emissions (capital firms) and local pollution (mines). Subsidises via: (a) no subsidy, (b) fines collected are redistributed in region of collection based on firms' fraction of total capacity, (c) Same as b, but a share of global north funds is reallocated to the south, (d) same as b but funds are multiplied by a government grant scheme\\
    Monetary & No Experiments & No Experiments & No Experiments & No Experiments & No Experiments\\
    Regulation & No Experiments & (1) Carbon-risk adjustment: firms rank in banks' pecking order becomes the average of their credit rank and their emissions rank. (2) Green public guarantees: government backs loans to green firms completely. (3) Green Basel-II: exclude loans to green firms from the credit mutiplier regulation, thus increasing total credit supply& No Experiments & No Experiments & No Experiments\\
    Other & Exogenous increases to fossil fuel prices mimicking a carbon tax / green investment subsidy, and decreases mimicking current fossil subsidies \citep[Done in][]{LampertiEtAl2020ClimateChangeGreen} & No experiments & Test carbon tax when probability of a electricity plant being renewable is 50\% exogenously (e.g. regulation driven) & No Experiments & No Experiments \\
\bottomrule
\end{xltabular}
\end{center}

\newpage
\bibliographystyle{agsm}
\bibliography{References_ABMIAMReview.bib}
\end{document}